\documentclass[useAMS,usenatbib]{mn2e}
\usepackage{graphicx}

\def\Eiso{{E_{\rm iso}}}
\def\Ep{{E_{\rm p}}}
\def\Eg{{E_{\gamma}}}
\def\Lp{{L_{\rm p}}}

\def\eiso{{E_{{\rm iso},\,53}}}
\def\eg{{E_{\gamma,\,51}}}
\def\kev{{{\rm keV}}}
\def\lp{{L_{{\rm p},\,52}}}

\title[Opening Angle Distribution of GRB]
{A possible observational evidence for $\theta^{-2}$
angular distribution of opening half-angle of GRB jets}
\author[D. Yonetoku et al.]
{Daisuke Yonetoku$^{1}$\thanks
{E-mail: yonetoku@astro.s.kanazawa-u.ac.jp (DY)} 
Ryo Yamazaki$^{2}$,
Takashi Nakamura$^{3}$,
and Toshio Murakami$^{1}$ \\
$^{1}$Department of Physics, Kanazawa University, 
Kakuma, Kanazawa, Ishikawa 920-1192, Japan\\
$^{2}$Department of Earth and Space Science,
Osaka University, Toyonaka 560-0043, Japan\\
$^{3}$Department of Physics, Kyoto University,
Kyoto 606-8502, Japan 
}
\begin{document}


\pagerange{\pageref{firstpage}--\pageref{lastpage}} \pubyear{2002}

\maketitle

\label{firstpage}

\begin{abstract}
We propose a method to estimate the pseudo jet 
opening half-angle 
of GRBs using the spectral peak energy 
($\Ep$)--peak luminosity 
relation (so called Yonetoku relation) as well as 
the $\Ep$--collimation-corrected $\gamma$-ray energy relation
(so called Ghirlanda relation). 
For  bursts with  known jet break times  and redshifts, 
we compared  the pseudo jet opening half-angle  
with the standard one and found that  the differences are 
within a factor 2. We apply the method to 689 long GRBS.
We found that the distribution function of the pseudo 
jet opening half-angle obeys $f(\theta_j)\propto\theta_j^{-2.2 \pm 0.2}$ with 
possible cutoffs for $\theta_j < 0.04$ and $\theta_j > 0.3$
although the log-normal fit is also possible.
$\theta^{-2}$ distribution is compatible with the structured jet model.
From the distribution function we found that 
the beaming correction
for the rate of GRBs is  $\sim 340$, which means 
$\sim 10^{-5}$~yr$^{-1}$~galaxy$^{-1}$ or only one in $10^2$~type 
Ib/c supernovae. We also found the evolution of the distribution 
function as a function of the redshift.
\end{abstract}

\begin{keywords}
gamma rays: bursts ---  gamma rays: observations
--- gamma rays: theory.
\end{keywords}

\section{Introduction}
Gamma-ray bursts (GRBs) arise from relativistic jets
\citep[e.g.,][]{m02,zm03}. Relativistic motion, with Lorentz factor of 
greater than  $\sim10^2$ is necessary in order to resolve the 
``Compactness problem''. On the other hand, the evidence of the jet 
collimation has been derived by the wave-length-independent achromatic 
break in the observed afterglow light curve. The jet opening half-angle has 
been actually measured in this context for bursts with successfully measured 
redshifts and jet break times. However, the small number of samples prevents 
us from discussing the statistical properties of the opening angle 
distribution. We have not yet  known the maximum and minimum value of 
the opening angle as well as the slope of the distribution function well
 \citep{frail01}.
These quantities are very important to argue the GRB rate, 
the energetics, the nature of  progenitors, and so on.

There is a  correlation between the rest-frame spectral peak energy 
$\Ep$ and the isotropic equivalent $\gamma$-ray energy $\Eiso$ of GRBs 
called the Amati relation 
\citep{a02,atteia2003,lamb2004,sakamoto04}.
Several authors have argued theoretical interpretations of the relation
\citep{yin04a,eichler04,ldg05}. While 
\citet{np04} argued against the Amati relation because
$\sim$40\% of 751 BATSE GRBs do not have the solution to
 the undetermined
redshift under the Amati relation and clear outliers such 
as GRB~980425 and GRB~031203 exist.
A similar argument against the Amati relation has also been made by
\citet{bp05}.
\citet{ggf05} argued against \citet{np04} as well as \citet{bp05}.
 They used 442 bright BATSE GRBs with good
statistics and pseudo redshift from the lag-luminosity relation
\citep{bnb04} and obtained the
Amati relation with the slightly different power law index 
and the larger scatter than the original one. They  found 
that the chance probability of the revised Amati relation is 
$2.1\times 10^{-65}$. 
Similar analysis can be also seen in \citet{bclb05}.

As for outliers, \citet{yyn03} argued that if the intrinsic $\Ep$
of GRB~980425 is 2--4~MeV similar to GRB990123
 and the viewing angle is 
$\sim \theta_j+10\gamma^{-1}$ with $\theta_j$ and $\gamma$ being
the jet opening half-angle and the gamma factor of the jet,
the observed $\Eiso$ and $\Ep$ of GRB~980425 can be reproduced.
If we locate GRB~980425 at the distance larger than 800~Mpc, the
event is not observed. 
Hence, GRB~980425-like events are not included
in the bright BATSE GRBs.
Similar arguments have been also done for GRB~031203 \citep{rgk04}.
 Note here that in the
classification of {\it HETE-2}, GRB~980425 and GRB~031203 are 
X-ray rich GRBs.

Recently, similar relations to the Amati one 
with the tighter correlation have been found.
\citet{yonetoku04} have shown that for 16 GRBs with known redshifts detected 
by BATSE and {\it BeppoSAX}, there is a tight positive correlation between 
$\Ep$ and the peak luminosity ($\Ep$--$\Lp$ relation).
The chance probability is extremely low 
$5.3\times10^{-9}$. 
\citet{ggfcb05}
called the  relation as the Yonetoku relation and checked
the validity of the relation using 
442 bright GRBs with the pseudo
 redshift and confirmed the relation with the same power law
index within the error in the original Yonetoku relation.
 They found that the chance probability of the Yonetoku 
relation is $1.6\times 10^{-69}$.
\citet{yonetoku04} used the Yonetoku relation as the 
redshift indicator for 689 GRB samples without known redshift, 
and derived the GRB formation rate.
On the other hand, for 15 GRBs detected by BATSE, 
{\it BeppoSAX} and {\it HETE-2} with measured 
redshift and opening half-angle $\theta_j$, \citet{ggl04} found that  $\Ep$ 
correlates with the collimation-corrected $\gamma$-ray energy $\Eg$.
We call the relation as the Ghirlanda relation.

The $E_p$-$L_p$ diagram of GRBs may correspond to
the Hertzsprung-Russell (HR) diagram 
of  stars.  Then the Yonetoku as well as 
the Ghirlanda relations
correspond to the main-sequence where stars cluster around
 a single curve determined by the mass of the star. 
This suggests  the existence of 
a certain parameter that controls  GRBs 
 like the  mass of the star in  the HR diagram.
In the HR diagram the outliers of the main sequence 
exist such as red giants and 
white dwarfs. We know the physical reasons for the existence of
these outliers so that they do not refute the main sequence 
relation. In the Yonetoku and the Ghirlanda 
relations also outliers
exist for which we will know the physical reasons in future
as suggested in \citep{yyn03,rgk04}.

In this letter, we show that the Yonetoku and the Ghirlanda
 relations
can be used to  estimate the opening half-angle of the relativistic jet of GRBs. Main advantage of our jet opening half-angle estimator is that the 
opening angle can be calculated only from the information of the prompt 
emission. This letter is organized as follows. In section~\ref{sec:angle}, 
using the Yonetoku and Ghirlanda relations, we derive an empirical 
formula to estimate the opening half-angle and discuss 
its validity. 
Using the estimator, we derive the distribution of opening half-angle of 
BATSE-triggered bursts in section~\ref{sec:distribution}.
Section~\ref{sec:disc} is devoted to discussions.


\section{Jet Opening Angle inferred from th Yonetoku and 
the Ghirlanda relations}
\label{sec:angle}

Usually  the jet opening half-angle is estimated only when both the redshift 
and the achromatic break time in the afterglow light curve are measured.
Under the simple assumption of a uniform ambient matter distribution
of number density $n_0$, the jet opening half-angle is estimated as
\begin{equation}
\theta_{j,\,{\rm break}}=0.12\
[t_{\rm jet,\, d}/(1+z)]^{3/8}(n_0\eta_\gamma/\eiso)^{1/8}~~,
\label{eq:break}
\end{equation}
where $\eiso=\Eiso/10^{53}$~erg, and
$t_{\rm jet,\, d}$ and $\eta_\gamma$ are the jet break time in days 
and the efficiency of the fireball in converting the energy in the ejecta 
into $\gamma$-rays, respectively  \citep{sari99}.
However continuous follow-up 
observations are required to measure the achromatic jet-break time,
and moreover these kinds of observations are realized only for the 
bright afterglows so that it is hard to measure the opening 
half-angle for large
amount of GRBs in this method.

We propose here a different method to estimate the opening 
half-angle using only 
informations of the prompt gamma-ray emission. 
Let us assume that the rest-frame spectral peak energy $\Ep$,
the peak luminosity $\Lp$, the isotropic equivalent $\gamma$-ray energy
$\Eiso$, the jet opening-half angle $\theta_j$,
and the collimation-corrected energy $\Eg$ satisfy 
the Yonetoku and the Ghirlanda relations as
\begin{equation}
\Ep = 2.1\times10^2\ \lp^{0.50}\ \kev~~,
\end{equation}
\begin{equation}
\Ep = 4.8\times10^2\ \eg^{0.71}\ \kev~~,
\end{equation}
\begin{equation}
\Eg=(1-\cos\theta_j)\Eiso~~,
\end{equation}
where $Q_x$ denotes $Q/10^x$ in cgs units \citep{yonetoku04, ggl04}.
From equations (2), (3) and (4), 
we have,
\begin{equation}
\Omega_j\equiv
1-\cos\theta_j = 0.30\ \lp^{0.50/0.71} E_{\rm iso, 51}^{-1}~~.
\label{eq:angle}
\end{equation}
From the observed $E_p$ and $L_p$, we can determine the 
redshift under the Yonetoku relation and the given cosmological
parameter as in \citep{yonetoku04}. Then $\Eiso$ can be 
computed using the observed fluence. 
Therefore, only from the information of the prompt gamma-ray
emission, we can estimate the jet opening half-angle 
for each GRB. 
This method has a strong advantage compared with the jet-break
measurement in afterglows since we can use the large number
of BATSE GRBs data.

\begin{figure}
\includegraphics[width=84mm]{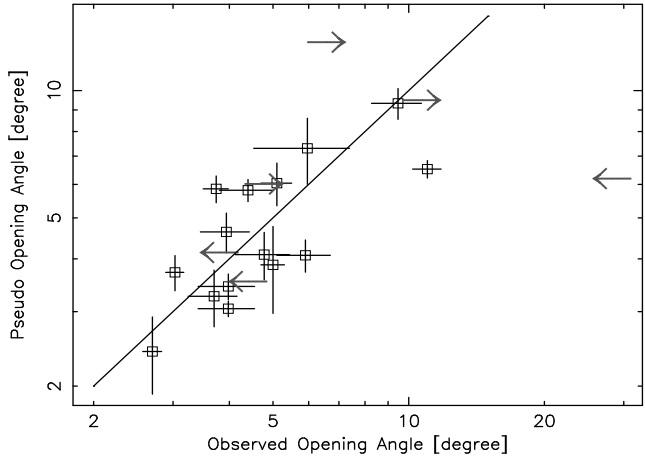}
 \vspace{0pt}
 \caption{
Comparison between jet opening half-angle $\theta_j$ estimated 
by Eq.~(\ref{eq:angle}) and $\theta_{j,\,{\rm break}}$ estimated by
Eq.~(\ref{eq:break}). The solid line is an equivalent line.
}
\label{fig:angle_compare}
\end{figure}

To show the validity of our method,
in figure~\ref{fig:angle_compare}, we compare $\theta_{j,\,{\rm break}}$ 
estimated by Eq.~(\ref{eq:break}) with $\theta_j$ by
Eq.~(\ref{eq:angle}) for GRBs with measured redshift and jet break time. 
It is found that there exists a positive correlation with the linear 
correlation coefficient including weighting factor is 0.760 with 13 
degree of freedom (excluding the lower and upper limit samples), 
which corresponds to the chance probability of $9.63\times 10^{-4}$. 
Their differences from the linear function (equivalent line) are within 
a factor of 2.  Although we had better use 
the word ``the pseudo jet opening half-angle'' for the half-angle from 
equation (4), we simply use the opening half-angle in this letter for 
convenience.

\section{Jet Opening Angle Distribution of BATSE Bursts}
\label{sec:distribution}

We used the same data set of 689 long GRBs published in \citep{yonetoku04}
to measure the distribution function, $f(\theta_j)$.
Here $\theta_j$ is the jet opening 
half-angle and $f(\theta_j)d\theta_j$ is the relative number
of the jet with the opening half-angle between 
$\theta_j$ and $\theta_j+d\theta_j$ .
In order to have a better signal-to-noise ratio 
in our analyses, 
we selected 605 GRBs with the flux greater than 
$1 \times 10^{-6}~{\rm erg~cm^{-2}s^{-1}}$. 
Having obtained the observed peak flux and the spectral indices as well 
as the $E_{p}$, the redshift is estimated using the Yonetoku relation
for each event \citep{yonetoku04}. Then, we can estimate $\Lp$ and $\Eiso$. 
Hence $\theta_j$ is also calculated by equation~(\ref{eq:angle}).
The empirical $\theta_j$ estimator shown in equation~(\ref{eq:angle})
 depends on 
two parameters; $L_{p}$ and $E_{iso}$. This fact means that the two different 
 types of selection effects are mixed. To avoid such doubly truncation 
effects, we manually set a truncation as the 
$E_{iso} > L_{p} \times 1~{\rm sec}$ 
at the rest frame of GRBs. In this case, both $L_{p}$ and $E_{iso}$ 
are limited only by the peak flux, so we can deal 
with the data as the simply truncated one.

\begin{figure}
\includegraphics[width=84mm]{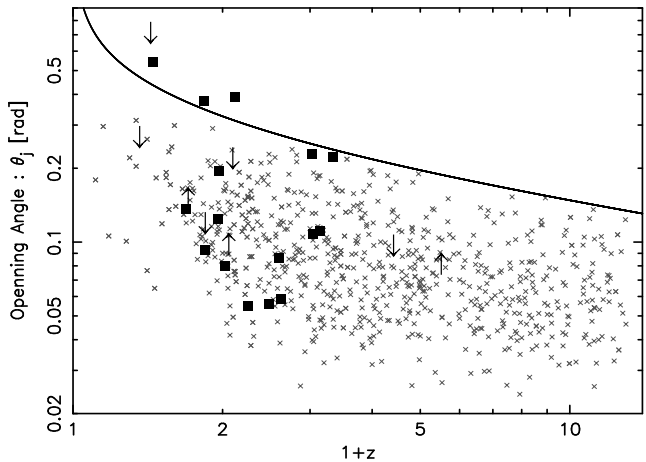}
 \vspace{0pt}
 \caption{
Distribution of the jet opening half-angle $\theta_j$ estimated 
by Eq.~(\ref{eq:angle}) vs. redshift derived from the $\Ep$--$\Lp$ 
relation. The solid line shows the truncation of the upper bound of 
$\theta_j$ that is caused by the flux limit of 
$F_{\rm lim}=2\times10^{-7}~{\rm ergs~cm^{-2}s^{-1}}$.
The cross points are the $\theta_{j}$ measured in the present work.
The opening angles measured from the jet-break time in the optical 
afterglow are also plot on the same figure as squares and arrows.
}
\label{fig:z_angle}
\end{figure}

Figure~\ref{fig:z_angle} shows the distribution of the derived
jet opening half-angle as a function of the redshift. The $\theta_{j}$ 
distributes within the range of $0.04 < \theta_j < 0.3$ radian.
Using the flux limited samples explained above, 
we can estimate the opening half-angle evolution and the 
true distribution of $\theta_j$ following 
the  detailed mathematical descriptions
\citep[e.g.][]{lynden, petro93, maloney, yonetoku04}.

\begin{figure}
\includegraphics[width=84mm]{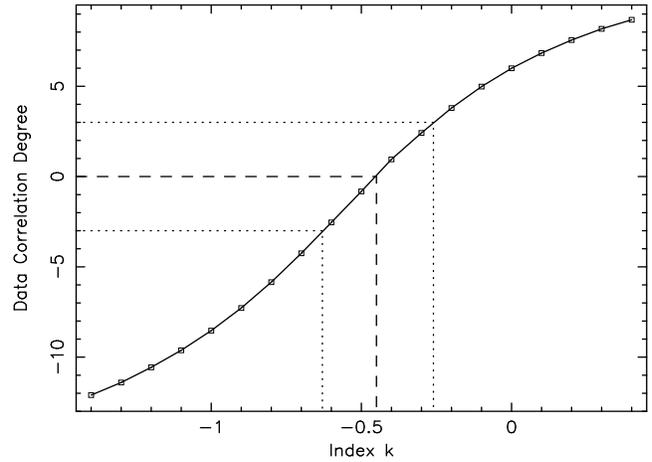}
 \vspace{0pt}
 \caption{
The best index of the jet opening angle evolution measured by the
$\tau$-statistical method. When we assume $\theta_{j}$ evolution as 
the function of $(1+z)^{k}$, $k = -0.45$ is the best value.
We also show 3~sigma upper and lower bound of the index as
$k = -0.26$ and $k = -0.63$, respectively. 
No evolution of $\theta_{j}$
was rejected about 6~$\sigma$ confidence.
}
\label{fig:index-tau}
\end{figure}

First, we estimate the opening half-angle evolution, which is the redshift 
dependence of $f(\theta_j)$, from the $(1+z, \theta_j)$ distribution shown 
in figure~\ref{fig:z_angle}. The data correlation degree in the flux 
limited samples are calculated by so called $\tau$-statistical method
which is very similar to the Kendell's $\tau$ statistics.
To refer to the previous works \citep{lloyd, yonetoku04}, we assume 
the functional form of   $f(\theta_j)$ evolution as $(1+z)^{k}$, and
calculate the data correlation degree for each $k$ value.
In figure~\ref{fig:index-tau}, we show the $(1+z, \theta_{j})$ correlation
as a function of index $k$. No $\theta_{j}$ evolution is rejected 
about 6 sigma confidence level. When we assume the case of $k = -0.45$,
$\theta_{j}$ becomes independent of the redshift. In other words,
$\theta_{j}$ evolution of $(1+z)^{-0.45}$ is hidden in the 
$(1+z, \theta_{j})$ plane.

\begin{figure}
\includegraphics[width=84mm]{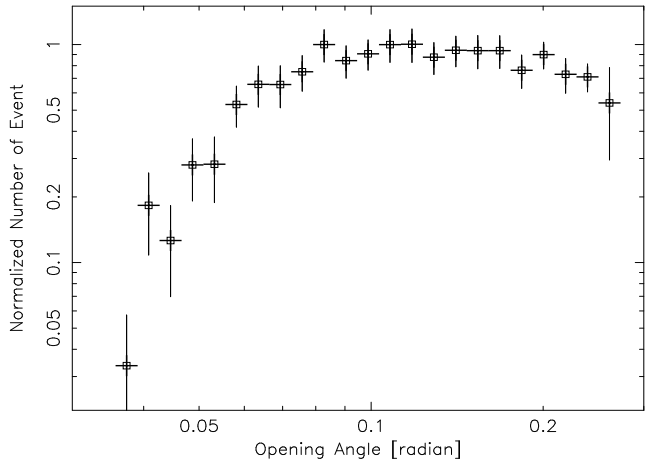}
 \vspace{0pt}
 \caption{
The differential $\theta_{j}'$ distribution at $z=1$ of the GRBs 
observed with BATSE detector.
}
\label{fig:result}
\end{figure}

Next, we derive the opening half-angle distribution. One possible method 
(non-parametric method) developed by \citet{lynden} is applied
in this analysis. We define $\theta_{j}' \equiv \theta_{j} /(1+z)^{-0.45}$
which is equivalent to the evolution-removed opening half-angle.
Then the observed data randomly distributes in the $(1+z, \theta_{j}')$
plane, so we can easily assume the missing data caused by the flux limit. 
Strictly speaking, for high-redshift GRBs (e.g., $z \ga 5$),
the observed fluence may be estimated as the lower value because there 
may be  missing photons behind the background level. On the other hand, 
the peak flux may be correctly estimated. Hence derived 
opening half-angles 
for high-$z$ events may be larger than the actual values. 
In order to avoid such confusion, we consider 430 events within 
the estimated redshifts of less than 4.5. Additional merit is that 
the k-correction factor can be neglected. The differential $\theta_{j}'$ 
distribution $f(\theta_{j}')$ which corresponds to one at
 $z = 1$, 
is shown in figure~\ref{fig:result}. If one would like to obtain the 
$\theta_{j}$ distribution at each redshift, it is roughly estimated as 
$f(\theta_{j}) = f(\theta_{j}') ((1+z)/2)^{-0.45}$.

\begin{figure}
\includegraphics[width=84mm]{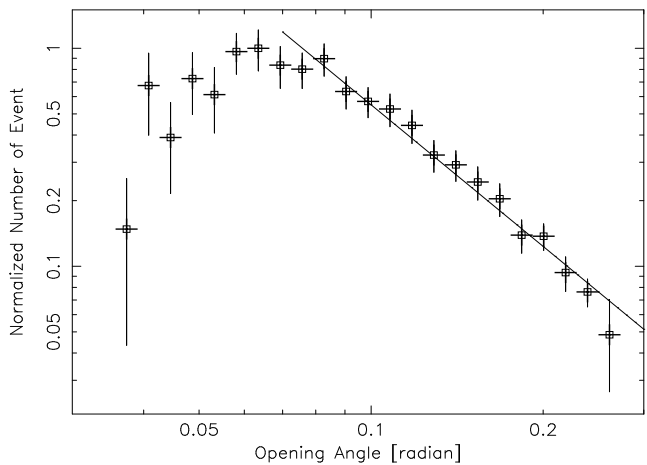}
 \vspace{0pt}
 \caption{
True opening angle distribution at $z=1$. When we adopt the power-law 
model for the region $\theta_{j} > 0.05$~radian, we obtain the best result
of $\theta_{j}'^{-2.2 \pm 0.2}$ with 90~\% confidence level.
}
\label{fig:result2}
\end{figure}

The opening half-angle distribution $f(\theta_{j}')$ is 
based on the 
number of event detected by BATSE. The chance probability to observe
narrowly collimated events are much lower than that of the wide 
opening half-angle events due to the geometrical effect. Therefore,
we have to take into account the correction of
$f(\theta_{j}')/\Omega_{j}$ when we argue the true probability 
density function. In figure~\ref{fig:result2}, we show the true
opening half-angle distribution. It can be fitted by the power-law form
in the range $\theta_{j}' > 0.05$~radian as $\theta_{j}'^{-2.2 \pm 0.2}$.
Clear cut off at $\sim 0.04$~rad can be seen.

\citet{ggf05} obtained the opening half-angle distribution
using the Ghirlanda relation and the Amati relation. The result
looks like a log-normal one. We also tried to fit a log-normal
one. We obtained acceptable results with the mean 
$\log \mu = -1.15$ (0.07~radian) and the standard deviation 
$\log \sigma = 0.24$ (0.04--0.12~radian). However the data distribution 
is asymmetric in the logarithmic horizontal scale as shown in 
Figure~\ref{fig:result2}.  To say  the power law 
distribution with cut-offs is better than the log-normal one,
 we have to confirm that the cut-off at  $\theta_{j} = 0.04$ exists. However  we can not conclude 
the existence because of the large errors in the small $\theta_{j}$ region.

From the distribution function we found the beaming correction
for the rate of GRBs $\sim 340$ at $z=1$ by computing 
$\langle\Omega^{-1}\rangle= 
\int f(\theta_j')\Omega^{-1}d\theta_j'/\int f(\theta_j')d\theta_j'$. 
This rate is about factor of 4 larger than that obtained by \citet{gpw04}
 using a different method, while \citet{frail01} obtained this correction 
$\sim$ 520 using the same method with smaller number of samples. Since the lower cut-off of the distribution $\theta_j' \sim 0.04$ are 
similar for both cases, the difference comes from the higher power
law index $\sim -4.54$ of \citet{frail01}. Since the number of 
GRBs is much smaller in their analysis, observational biases of the
 smaller samples of larger opening half-angle causes the
steeper power law. Adopting the local GRB rate of 
$\sim 0.5$~Gpc$^{-3}$yr$^{-1}$ \citep{schmidt01}, 
we obtain the true rate of GRB as 
$\sim 170$~Gpc$^{-3}$yr$^{-1}$ which means
$\sim 10^{-5}$~yr$^{-1}$galaxy$^{-1}$ with the number density
of the galaxy being $\sim 10^{-2}$~galaxy~Mpc$^{-3}$. 
Since the rate of type Ib/c supernovae is
$\sim10^{-3}$~yr$^{-1}$galaxy$^{-1}$ \citep{cbt03}, only one in
$10^2$~type~Ib/c supernovae becomes GRB if GRB is the peculiar type~Ib/c supernova.

\section{Discussions}
\label{sec:disc}

We present an empirical
opening half-angle  estimator that is
inferred from the Yonetoku and the Ghirlanda relations.
Our method requires only  the data of prompt emission,
which is different from the standard method using the
break in the afterglow light curve and the redshift.
Using the empirical opening angle estimator,
we have derived the opening angle distribution of GRBs.
The distribution can be fitted by the power-law form
in the range $\theta_j >0.07$~rad as $\theta_{j}^{-2.2\pm0.2}$.
The cut off at $\sim0.04$~rad can be seen.
In the uniform jet model of GRBs, this means that the 
distribution function happens to be $\sim \theta^{-2}$.

The other possibility is the structured jet model.
In the original version of the universal structured jet 
model \citep{rossi2002,zm02}, the energy per unit solid angle 
is in proportion to $\sim \theta^{-2}$.  The lower and upper 
angle cutoff exist
and the jet structure is essentially the same between the lower 
and the upper cutoff, which is the origin of the name 
``structured'' jet.
The viewing angle corresponds 
to the jet opening angle in the structured jet model so that
we need to argue  observationally what will be the opening angle
distribution of the structured jet \citep{psf03}.
If all bursts were observable, the distribution would be uniform 
per unit solid angle and $f(\theta)\propto\theta$. 
However, $\Eiso$ for the
smaller viewing angle is brighter by a factor of $\theta^{-2}$
so that the maximum observable distance is larger by a
factor of $\theta^{-1}$ which contains a volume larger than
$\theta^{-3}$. Then we have $f(\theta)\propto\theta^{-2}$
which is compatible with the result in the present paper.
Especially our Fig.~5 looks like Fig.~3 of \citet{psf03}
 with appropriate parameters.

The evolution effect found in the present
paper means that the jet opening half-angle becomes narrower
for larger redshift. One possible qualitative explanation is
that this is due to the metal dependence of the progenitors.
Since the metalicity  of the star decreases
as a function of the redshift, it is expected that
the mass loss of the stars decreases as a function of the
redshift if the mass loss is derived by the line absorption of
photons in the atmosphere of the stars. This suggests that
for high redshift progenitors of GRBs, the mass of  
the envelope is larger so that only the sharper jet can
punch a hole in the envelope of the progenitor star.   

In this paper, when we calculate the true opening half-angle distribution,
we implicitly assumed that the jet emission can be seen
only when the jet is seen on-axis.
In reality,
effects of off-axis viewing GRBs might be important
\citep{yin02,yin03b,yin04a}.
When the contributions of off-axis emission is considered,
 true distribution may be modified.
However, the beaming correction from off-axis effects is important only
for $z\la1.5$ because of the relativistic beaming effect
\citep{yin04a}.

\section*{Acknowledgments}

This work was supported in part by
a Grant-in-Aid for for the 21st Century COE
``Center for Diversity and Universality in Physics''
and also supported by Grant-in-Aid for Scientific Research
of the Japanese Ministry of Education, Culture, Sports, Science
and Technology, No.15740149 (DY), No.05008 (RY),
No.14047212 (TN), and No.14204024 (TN).

%



\label{lastpage}

\end{document}